\documentclass[runningheads]{llncs}
\usepackage{nopageno}
\usepackage{amssymb}
\usepackage{graphicx}

\usepackage{url}
\urldef{\mailsa}\path|{iberges003, jesus.bermudez, alfredo, a.illarramendi}@ehu.es|
\newcommand{\keywords}[1]{\par\addvspace\baselineskip
\noindent\keywordname\enspace\ignorespaces#1}

\begin{document}

\mainmatter  

\title{Semantic Web Technology for Agent Communication Protocols}

\titlerunning{Semantic Web Technology for Agent Communication Protocols}

%
%
\author{Idoia Berges\thanks{The work of Idoia Berges is supported 
by a grant of the Basque Government.}\and Jes\'us Berm\'udez\and Alfredo Go\~ni \and Arantza Illarramendi\thanks{All authors are members of the Interoperable DataBases Group. This work is also supported by the University of
the Basque Country, Diputaci\'on Foral de Gipuzkoa (cosupported by the European Social
Fund) and the Spanish Ministry of Education and Science TIN2007-68091-C02-01.}}
\authorrunning{Semantic Web Technology for Agent Communication Protocols}

\institute{University of the Basque Country\\
\mailsa\\
\url{http://siul02.si.ehu.es}}

%
%

\toctitle{Lecture Notes in Computer Science}
\tocauthor{Authors' Instructions}
\maketitle

\begin{abstract}

One relevant aspect in the development of the Semantic Web framework is the achievement of a real inter-agents communication capability at the semantic level. The agents should be able to communicate and understand each other using standard communication protocols freely, that is, without needing a laborious a priori preparation, before the communication takes place.

For that setting we present in this paper a proposal that promotes to describe standard communication protocols using Semantic Web technology (specifically, OWL-DL and SWRL). Those protocols are constituted by communication acts. In our proposal those communication acts are described as terms that belong to a communication acts ontology, that we have developed, called \textsc{CommOnt}. The intended semantics associated to the communication acts in the ontology is expressed through social commitments that are formalized as fluents in the Event Calculus.

In summary, OWL-DL reasoners and rule engines help in our proposal for reasoning 
about protocols. We define some comparison relationships (dealing with notions of 
equivalence and specialization) between protocols used by agents from different systems.

\end{abstract}

\section{Introduction}

In the scenario that promotes the emergent Web, administrators of existing Information Systems, that belong to nodes distributed along the Internet network, are encouraged to provide the functionalities of those systems through agents that represent them or through Web Services.
The underlying idea is to get a real interoperation  among those Information Systems in order to enlarge the benefits that users can get from the Web by increasing the machine processable tasks.

Although agent technology and Web Services technology have been developed in a separate way, there exists a recent
work of several members from both  communities trying to consolidate
their approaches into a common specification describing how
to seamlessly interconnect FIPA compliant agent systems \cite{FIPA-ACL}
with W3C compliant Web Services.
The purpose of specifying an infrastructure for integrating
these two technologies is to provide a common means of
allowing each to discover and invoke instances of the other 
\cite{GLMS07}. Considering the previous approach, in the rest of this paper we will only concentrate on inter-agent communication aspects. 

Communication among agents is in general based on the interchange of communication acts. However, different Information Systems have incorporated different classes of communication acts as their Agent Communication Language  (\textsc{acl}) to the point that they do not understand each other. Moreover, protocols play a relevant role in agents communication. A protocol specifies the rules of interaction between agents by restricting the range of allowed follow-up communication acts for each agent at any stage during a communicative interaction. It is widely recognized the interest of using standard communication protocols.

We advocate so that the administrators of the Information Systems proceed in the following way. When they wish to implement the agents that will represent their systems, they first select, from a repository of standard protocols (there can exist one or more repositories), those protocols that fulfill the goals of their agents. Sometimes a single protocol will be sufficient and other times it will be necessary to design a protocol as a composition of some other protocols. Next, they can customize the selected protocols before they incorporate them to the agents. In that setting, when agents of different Information Systems want to interoperate it will be relevant to reason about the protocols embedded in the agents in order to discover relationships such as equivalence or restriction between them. Moreover, once those relationships are discovered both agents can use the same protocol by 
replacing dynamically in one agent the protocol supported by the other. Finally, 
in our opinion it will be desirable to use a formal language to represent the protocols.

In this paper we present a proposal that promotes to describe standard communication protocols using Semantic Web Technology (OWL-DL and SWRL). In addition, communication acts that take part of the protocols  are described as terms that belong to a communication acts ontology, that we have developed, called \textsc{CommOnt} (see more details about the ontology 
in~\cite{IS2007}). 
The use of that ontology 
favours on the one hand, the explicit representation of the meaning of the 
communication acts and on the other hand, the customization of existing standard protocols by allowing the use of particular communication acts that can be defined as specializations of existing standard communication acts. 

Terms of the \textsc{CommOnt} ontology are described using OWL-DL and 
we have adopted the so called \textit{social approach} \cite{Singh98,Singh00}
for expressing the  
intended semantics of the communication acts included in the protocols.
According to the social approach, when agents interact they become 
involved in social commitments or obligations to each other. Those commitments are public, and therefore they are 
suitable for an objective and verifiable semantics of agent interaction. 
Social commitments can be considered as \textit{fluents} in the Event Calculus, which is a logic-based 
formalism for representing actions and their effects. 
Fluents are propositions that hold during time intervals.
A formula in the Event Calculus is associated to a communication act for describing 
its social effects. 
The set of fluents that hold at a  moment describes a state of the interaction.
DL axioms and Event Calculus formulae apply to different facets of communication acts. 
DL axioms describe 
static features and are principally used for communication act interpretation purposes. Event Calculus formulae 
describe dynamic features, namely the social effects of communication acts, and are principally used for communication act 
operational contexts such as supervising conversations.

 In summary the main contributions of the proposal presented in this paper are:
\begin{itemize}
\item It favours a flexible interoperation among agents of different systems by using standard communication protocols described through tools promoted by the W3C.
\item It facilitates the customization of those standard communication protocols allowing to use communication acts in the protocols that belong to specific
\textsc{acl} of Information Systems. The particular communication acts are described in an ontology.
\item It provides a basis to reason about relationships between two protocols in such a way that the following relations can be discovered: equivalence or restriction (and also considering a notion of specialization). Moreover, notice that our approach allows to get protocols classification in terms of the 
intended semantics of communication acts that appear in the protocols.
\item It allows  modeling the communication among agents without regarding only to the lower level operational details of how communication acts are interchanged but taking also into account the meaning of those acts.

\end{itemize}

The rest of the paper is organized as follows: Section~\ref{pilares} provides 
background on the communication ontology, that contains terms corresponding to 
communication acts that appear in the protocols, and on the semantics associated to those acts. Section~\ref{description} explains how protocols are described using Semantic Web Technology and presents the definitions of the relationships considered between protocols. Section~\ref{trabajos} discusses different related works, and conclusions appear in the last section.

\section {Two basic supports for the proposal: the \textsc{CommOnt} Ontology and the representation of the semantics of communication acts}\label{pilares}
\noindent Among the different models proposed for representing protocols one which stands out is that of State Transition Systems (STS).
\begin{definition}
A \textit{State Transition System} is a tuple ($S$, $s_{0}$, $L$, $T$, $F$), where 
$S$ is a finite set of states, $s_{0} \in S$ is an initial state, $L$ is a finite set 
of labels, $T\subseteq S\times L\times S$ is a set of transitions and $F\subseteq S$ 
is a set of final states. 
\end{definition}
In our proposal we use STS where transitions are labeled with communication act classes  
described in a communication acts ontology called \textsc{CommOnt}. 
That is to say, the set of labels $L$ is a set of class names taken from that ontology. 
Moreover, as mentioned before, the 
intended semantics associated to the communication acts in the ontology is expressed through 
predicates in the Event Calculus that initiate or terminate fluents. In our case, each state 
is associated to the set of fluents that holds at that moment.

In the following two subsections we present the main features of  the 
\textsc{CommOnt} ontology  and of the intended semantics associated to communication acts, 
respectively.

\subsection{{Main features of the \textsc{CommOnt} Ontology}}

The goal of the \textsc{CommOnt} ontology is to favour the interoperation 
among agents belonging to
different Information Systems.
The leading categories of that ontology are: first,
\textit{communication acts} that are used for interaction by
\textit{actors} and that have different purposes and deal with
different kinds of contents; and second, \textit{contents} that are
the sentences included in the communication acts.

The main design criteria adopted for the communication acts category
of the \textsc{CommOnt} ontology is to follow the \textit{speech
acts} theory \cite{Austin62}, a linguistic theory that is
recognized as the principal source of inspiration for designing the
most familiar standard agent communication languages.  Following that
theory every communication act is the sender's expression of an
attitude toward some possibly complex proposition. A sender performs a
communication act which is expressed by a coded message and is
directed to a receiver. Therefore, a communication act has two main
components.  First, the attitude of the sender which is called the
\textit{illocutionary force} (\textit{F}), that expresses social
interactions such as informing, requesting or promising, among
others. And second, the \textit{propositional content} (\textit{p})
which is the subject of what the attitude is about.
In \textsc{CommOnt}  this \textit{F(p)} framework is followed, and 
different kinds of illocutionary forces and contents
leading to different classes of communication acts are supported.  More
specifically, specializations of illocutionary forces
that facilitate the absorption of aspects of the content into the
illocutionary force are considered.

\textsc{CommOnt} is divided into three interrelated layers: 
\textit{upper}, \textit{standards} and \textit{applications}, that group communication acts at
different levels of abstraction. 
Classes of the \textsc{CommOnt} ontology are described using the Web
Ontology Language \textsc{ OWL-DL}. Therefore, communication acts among
agents that commit to \textsc{CommOnt} have an abstract
representation as individuals of a shared universal class of
communication acts.

In the upper layer, according to Searle's speech acts theory, five upper classes of communication acts 
corresponding to \textit{Assertives}, \textit{Directives}, \textit{Commissives}, \textit{Expressives} 
and \textit{Declaratives} are specified. But also the top class \texttt{CommunicationAct}\footnote{\texttt{This type} 
style refers to terms specified in the ontology.} is defined, which represents the universal class of 
communication acts. Every particular communication act is an individual of this class. 
In \textsc{CommOnt}, components of a class are represented by properties. 
 The most immediate properties of \texttt{CommunicationAct} are the content 
and the actors who send and receive the communication act. There are some other properties related
to the context of a communication act such as the conversation in which it is inserted or a link to the 
domain ontology that includes the terms used in the content.

A standards layer extends the upper layer of the ontology with specific terms that represent
classes of communication acts of general purpose agent communication languages,
like those from \textsc{KQML} or \textsc{FIPA-ACL}.
Although the semantic framework of those agent communication languages
may differ from the semantic framework adopted in \textsc{CommOnt}, in our opinion 
enough basic concepts and principles are shared to such an extent that a commitment to
ontological relationships can be undertaken in the context of the interoperation of
Information Systems.

With respect to \textsc{FIPA-ACL}, we can observe that it proposes four 
primitive communicative acts \cite{FIPA-ACL}: \textit{Confirm}, \textit{Disconfirm},
\textit{Inform} and \textit{Request}. 
The terms \texttt{FIPA-Confirm}, \texttt{FIPA-Disconfirm}, \texttt{FIPA-Inform} and 
\texttt{FIPA-Request} are used to respectively represent them as classes in \textsc{CommOnt}. Furthermore, the rest of the {\textsc FIPA} communicative acts are derived from those mentioned four primitives. Analogously, communication acts from {\textsc KQML} can be analyzed and the corresponding terms
in \textsc{CommOnt}  specified. It is of vital relevance for the interoperability aim to be able of
specifying ontological relationships among classes of different standards.
 
Finally,
it is often the case that every single Information System uses a limited
collection of communication acts that constitute its particular agent communication language. 
The applications layer reflects the terms describing communication acts used in such 
particular Information Systems.
The applications layer of the \textsc{CommOnt} ontology provides a framework for the 
description of the nuances of such communication acts. 
Some of those communication acts can be defined as particularizations
of existing classes in the standards layer and maybe some others as particularizations of
upper layer classes.
Interoperation between agents of two systems using different kinds of communication acts will proceed 
through these upper and standard layer classes. 

Following we show some axioms in the \textsc{CommOnt} ontology. 
For the presentation we prefer a logic notation instead of the more verbose \textsc{owl/\textsc{xml}} syntax.

\scriptsize
\begin{eqnarray*}
\texttt{CommunicationAct} & \sqsubseteq & \texttt{=1 hasSender.Actor} \sqcap \forall \texttt{hasReceiver.Actor} \sqcap \\ & & \forall \texttt{hasContent.Content} \\
\texttt{Request} & \sqsubseteq & \texttt{Directive} \sqcap \exists \texttt{hasContent.Command}\\
\texttt{Accept} & \sqsubseteq & \texttt{Declarative}\\
\texttt{Responsive} & \sqsubseteq & \texttt{Assertive} \sqcap \exists \texttt{inReplyTo.Request}
\end {eqnarray*}

\normalsize
\subsection{Semantics associated to Communication Acts}\label{semantics}
Formal semantics based on mental concepts such as \textit{beliefs}, \textit{desires} and \textit{intentions} 
have been developed for specifying the semantics of communication acts. However, they have been 
criticized on their approach \cite{Singh98} as well as on their analytical difficulties 
\cite{Wooldridge00}. We have adopted the so called social approach \cite{Singh00,Venkatraman99,Fornara02} 
to express the intended semantics of communication acts described in the \textsc{CommOnt} ontology. According to 
the social approach, when agents interact they become involved in social commitments or obligations to 
each other. 

\begin{definition}
A \textit{base-level commitment} \textsf{C}(\textit{x, y, p}) is a ternary relation representing 
a commitment made by \textit{x} (the \textit{debtor}) to \textit{y} (the \textit{creditor}) to bring 
about a certain proposition \textit{p}.
\end{definition} 

Sometimes an agent accepts a commitment only if a certain condition holds or, interestingly, only 
when a certain commitment is made by another agent. This is called a conditional commitment.

\begin{definition}
A \textit{conditional commitment} \textsf{CC}(\textit{x, y, p, q}) is a quaternary relation 
representing that if the condition \textit{p} is brought out, \textit{x} will be committed to 
\textit{y} to bring about the proposition \textit{q}.
\end{definition}

Moreover, the formalism we use for reasoning about commitments is based on the Event Calculus. 
The basic ontology of the Event Calculus comprises \textit{actions}, \textit{fluents} and \textit{time points}. It also 
includes predicates for saying what happens when (\textit{Happens}), for describing the initial situation (\textit{Initially}), for describing 
the effects of actions (\textit{Initiates} and \textit{Terminates}), and for saying what fluents hold at what times (\textit{HoldsAt}). 
See~\cite{Shanahan99} for more explanations.

Commitments (base-level and conditional) can be considered fluents, and semantics of communication acts can be expressed with predicates. For example:

\begin{itemize}
\item 
\textit{Initiates(Request(s,r,P)}, \textsf{CC}\textit{(r, s, accept(r,s,P), P), t)} \\ 
A Request from \textit{s} to \textit{r} produces the effect of generating a conditional commitment expressing that if the receiver \textit{r} accepts the demand, it will be commited to the proposition in the content of the communication act.
\item 
\textit{Initiates(Accept(s,r,P), accept(s,r,P), t)}\\
The sending of an Accept produces the effect of generating the accept fluent.
\end{itemize}

Furthermore, some rules are needed to capture the dynamics of commitments. Commitments 
are a sort of fluents typically put in force by communication acts and that become inoperative after the 
appearance of other fluents. In the following rules \textit{e(x)} represents an event caused by 
\textit{x}. The first rule declares that when a debtor of a commitment that is in force causes an event that 
initiates the proposition committed, the commitment ceases to hold.

\textsc{Rule 1:}  
\textit{HoldsAt(}\textsf{C}\textit{(x, y, p), t)} $\wedge$ \textit{Happens(e(x), t)} $\wedge$ \textit{Initiates(e(x), p, t)} $\rightarrow$ 
\textit{Terminates(e(x),}\textsf{C}\textit{(x, y, p), t)}.

The second rule declares that a conditional commitment that is in force disappears and generates a base-level 
commitment when the announced condition is brought out by the creditor.

\textsc{Rule 2:}  
\textit{HoldsAt(}\textsf{CC}\textit{(x, y, c, p), t)} $\wedge$ \textit{Happens(e(y), t)} $\wedge$ \textit{Initiates(e(y), c, t)} $\rightarrow$
\textit{Initiates(e(y),}\textsf{C}\textit{(x, y, p), t)} $\wedge$ 
\textit{Terminates(e(y),}\textsf{CC}\textit{(x, y, c, p), t)}.
%

Following we state some predicates that describe the semantics asociated to some of the communication acts of the upper level of the \textsc{CommOnt} ontology. This semantics is determined by the fluents that are initiated or terminated as a result of the sending of a message between agents.

\begin{itemize}
\item \textit{Initiates(Assertive(s,r,P), {P}, t)}
\item \textit{Initiates(Commissive(s,r,C,P), }\textsf{CC}\textit{(s,r,C,P), t)}
\item \textit{Initiates(Responsive(s,r,P, RA), \textit{P}, t)}\\
\textit{Terminates(Responsive(s,r,P, RA),}\textsf{C}\textit{(s,r,RA), t)}
\end{itemize}
Effects of these predicates can be encoded with SWRL rules. 
For instance, the predicate 
\textit{Initiates(Request(s, r, P), }\textsf{CC}\textit{(r, s, accept(r, s, P), P), t)} can be 
encoded as follows:

\textit{Request(x)} $\wedge$ \textit{hasSender(x,s)} $\wedge$ \textit{hasReceiver(x,r)} $\wedge$ \textit{hasContent(x,p)} $\wedge$ \textit{hasCommit(x,c)} $\wedge$ \textit{isConditionedTo(c,a)} $\wedge$ \textit{atTime(x, t)}  $\rightarrow$ \textit{initiates(x,c)} $\wedge$ \textit{hasDebtor(c,r)} $\wedge$ \textit{hasCreditor(c,s)} $\wedge$ \textit{hascondition(c,p)} $\wedge$ \textit{Acceptance(a)} $\wedge$ \textit{hasSignatory(a,r)} $\wedge$ \textit{hasAddressee(a,s)} $\wedge$ \textit{hasObject(a,p)} $\wedge$  \textit{atTime(c, t)}

\section{Protocol Description}\label{description} 

As mentioned in the introduction, our proposal promotes to describe standard protocols 
using Semantic Web technology. We use STS as models of protocols. More specifically, 
we restrict to deterministic STS (i.e. if ($s$, $l$, $s'$) $\in T$ and 
($s$, $l$, $s''$) $\in T$ then $s'=s''$).
In order to represent protocols using OWL-DL, we have defined 
five different classes: \texttt{Protocol}, \texttt{State}, \texttt{Transition}, 
\texttt{Fluent} and \texttt{Commitment}, which respectively represent protocols, states, transitions in protocols, fluents and commitments associated to states. 

We model those class descriptions with the following guidelines:
A state has fluents that hold in that point and  transitions that go out of it. 
A transition is labelled by the communication act that is sent and is associated to 
the state that is reached with that transition. 
A fluent has a time stamp that signals the moment it was initiated. 
An actual conversation following a protocol is an individual of the class \texttt{Protocol}.
Following are some of the ontology axioms:

\scriptsize
\begin{eqnarray*}
  \texttt{Protocol}& \equiv & \exists \texttt{hasInitialState.State} \sqcap \\
  & & \forall \texttt{hasInitialState.State}   \\
  \texttt{State}& \equiv & \forall \texttt{hasTransition.Transition} \sqcap \\
  & & \exists \texttt{hasFluent.Fluent} \sqcap \\
  & & \forall \texttt{hasFluent.Fluent}\\
  \texttt{Transition}& \equiv & \texttt{=1 hasCommAct.CommunicationAct} \sqcap \\
  & & \texttt{=1.hasNextState.State}\\
  \texttt{FinalState}& \sqsubseteq & \texttt{State} \sqcap \\
  & & \forall \texttt{hasFluent.(Fluent} \sqcap \neg \texttt{Commitment)}\\
  \texttt{Fluent}&\sqsubseteq& \texttt{ =1 atTime}\\
 \texttt{Commitment}&\sqsubseteq& \texttt{Fluent} \sqcap  \texttt{ =1 hasDebtor.Actor} \sqcap\\
  & &  \texttt{=1 hasCreditor.Actor} \sqcap\\
  & &   \texttt{=1 hasCondition.Fluent}\\
  \texttt{ConditionalCommitment}&\sqsubseteq& \texttt{Fluent} \sqcap  \texttt{ =1 hasDebtor.Actor} \sqcap \\
  & &  \texttt{=1 hasCreditor.Actor} \sqcap \\
  & &   \texttt{=1 hasCondition.Fluent} \sqcap \\  
  & &   \texttt{=1 isConditionedTo.Fluent} \\
\end{eqnarray*}


\normalsize

The OWL-DL description of protocols reflects their static features and can be used to discover structural relationships between protocols. For instance, in Fig.~\ref{aplicarReglas} 
we show a simple protocol where agent A asks for time to agent B. The protocol description 
appears in the following:
\begin{figure}
	\centering
	\includegraphics[width=3.5in]{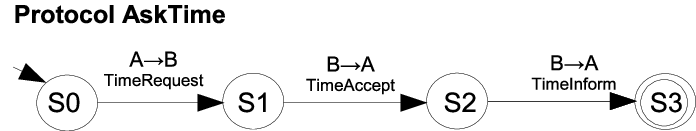}
	\caption{Protocol AskTime.} \label{aplicarReglas}
\end{figure}
\normalsize

\scriptsize
\begin{eqnarray*}
\texttt{Asktime} & \equiv & \texttt{Protocol} \sqcap \exists \texttt{hasInitialState.S0}\\
\texttt{S0} & \equiv & \texttt{State} \sqcap \exists \texttt{hasTransition.T01} \sqcap \exists \texttt{hasFluent.F0}\\
\texttt{S1} & \equiv & \texttt{State} \sqcap \exists \texttt{hasTransition.T12} \sqcap \exists \texttt{hasFluent.F1}\\
\texttt{S2} & \equiv & \texttt{State} \sqcap \exists \texttt{hasTransition.T23} \sqcap \exists \texttt{hasFluent.F2}\\
\texttt{S3} & \equiv & \texttt{FinalState} \sqcap \exists \texttt{hasFluent.F3}\\
\texttt{T01} & \equiv & \texttt{Transition} \sqcap \exists \texttt{hasCommAct.TimeRequest} \sqcap \exists \texttt{hasNextState.S1}\\
\texttt{T12} & \equiv & \texttt{Transition} \sqcap \exists \texttt{hasCommAct.TimeAccept} \sqcap \exists \texttt{hasNextState.S2}\\
\texttt{T23} & \equiv & \texttt{Transition} \sqcap \exists \texttt{hasCommAct.TimeInform} \sqcap \exists \texttt{hasNextState.S3}\\
\texttt{TimeRequest} & \equiv & \texttt{Request} \sqcap \texttt{=1 hasContent.TimeReq}\\
\texttt{TimeAccept} & \equiv & \texttt{Accept} \sqcap \texttt{=1 hasContent.TimeReq}\\
\texttt{TimeInform} & \equiv & \texttt{Responsive} \sqcap \texttt{=1 hasContent.TimeInfo} \sqcap \texttt{=1 inReplyTo.TimeRequest}
\end{eqnarray*}
\normalsize

 However, dealing only with structural relationships is too rigid if a flexible interoperation among agents that use different standard protocols is promoted. 
For that reason, we propose to consider what we call \textit{protocol traces}.
\begin{definition}
A \textit{protocol trace} is a sequence of time stamped fluents sorted in 
increasing order of time stamp.
\end{definition}
Notice that protocol traces are defined in terms of the semantics of communication acts, 
not in terms of the communication acts themselves; in contrast with many other related 
works (see section \ref{trabajos}) that consider messages as atomic acts without considering 
their content, neither their semantics.

During a simulation of a protocol run we apply the SWRL rules that encode the semantics 
of the communication acts (see section~\ref{semantics}) appearing in the run.
Then, we can consider the sorted set of time stamped fluents that hold at a final 
state of the protocol. 
That set represents the effects of the protocol run. 
Following we show an example of the application of the rules to 
a run of protocol AskTime in Fig.~\ref{aplicarReglas}.

\begin{figure}
	\centering
	\includegraphics[width=2.5in]{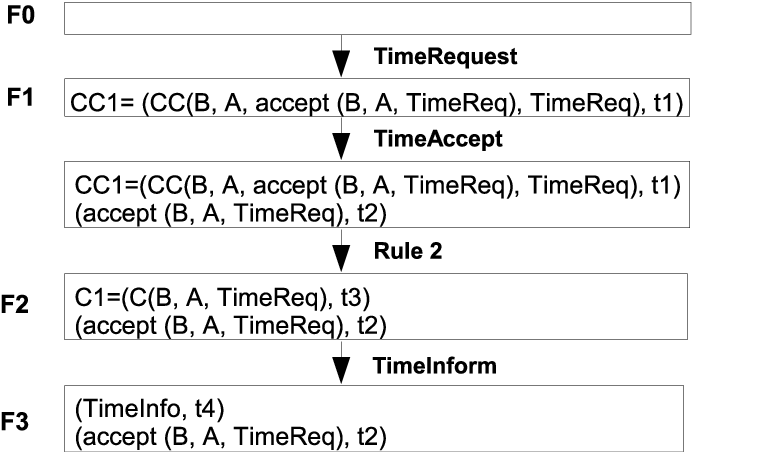}
	\caption{Protocol fluents} \label{fluentes}
\end{figure}

In Fig.~\ref{fluentes} we show which are the fluents associated to the states of the protocol and how they vary as a consequence of the communication acts that are sent and the rules described in section~\ref{semantics}.
\normalsize
We depart from a situation where the set of fluents is empty (F0). When the \texttt{TimeRequest} message is sent, due to the predicate \textit{Initiates(Request(s, r, P), \textsf{CC}(\textit{r, s, accept(r,s,P), P}), t)} the conditional commitment \texttt{CC1} 
is initiated, which states that if agent B accepts to give information about the time, then 
it will be committed to do so; $t_{1}$ is the time stamp associated. 
By convention we sort time stamps by their subindexes, that is: $t_{i}<t_{j}$ if $i<j$. 
 Then agent B agrees to respond by sending the \texttt{TimeAccept} message, and due to the predicate \textit{Initiates(Accept(s,r,P), accept(s,r,P), t)}, the fluent 
\textit{accept(B, A, TimeReq)} is initiated at time $t_{2}$. At this point, Rule 2 (see section~\ref{semantics}) can be applied, so \texttt{CC1} is terminated and the base 
commitment \texttt{C1} is initiated at time $t_{3}$. Finally, agent B sends the \texttt{TimeInform} message, and because of the predicates 
\textit{Initiates(Responsive(s,r,P, RA), P, t)} and \textit{Terminates(Responsive(s,r,P, RA), \textsf{C}(\textit{s,r,RA}), t)}, \texttt{C1} is terminated and a new fluent, \textit{TimeInfo}, is initiated at time $t_{4}$. 
So, at this point we can say that the fluents that hold at the final state of the protocol 
are (\textit{accept(B, A, TimeReq)}, $t_{2}$)  and (\textit{TimeInfo}, $t_{4}$). 

Then, we say that the protocol trace 
[(\textit{accept(B, A, TimeReq)}, $t_{2}$), (\textit{TimeInfo}, $t_{4}$)] 
is \textit{generated} by the protocol. We denote $\mathcal{T}(A)$ to the set
of all protocol traces generated by a protocol A.

Now, we proceed with the definitions of relationships between protocols we are 
considering. Our relationships are not structure-based but effect-based. 
Intuitively, two protocols are equivalent if the same effects take place 
in the same relative order. Runs of a protocol are made up of communication acts, 
and fluents are the effects they leave. 

\begin{definition}
Protocol A is \textit{equivalent} to protocol B if $\mathcal{T}(A) = \mathcal{T}(B)$.
\end{definition}
Sometimes, a protocol is defined by restrictions on the allowable communication acts at 
some states of a more general protocol. In those situations the application of those restrictions is reflected in the corresponding effects.
\begin{definition}
Protocol A is a \textit{restriction} of protocol B if $\mathcal{T}(A) \subset  \mathcal{T}(B)$.
\end{definition}
Protocols for specific Information Systems may use specialized communication acts. 
Specialization can also be applied also to domain actions that can be represented by specialized 
fluents.

\begin{definition}
A protocol trace $t$ is a \textit{specialization} of a protocol trace $s$,  
written $t{\ll}s$, if $\forall i.\ t(i)\sqsubseteq s(i)$ in an ontology of fluents.
\end{definition}
\begin{definition}
Protocol A is a \textit{specialized-equivalent} of protocol B if 
$\forall t \in \mathcal{T}(A).$ $\exists s \in \mathcal{T}(B).\ t{\ll}s$ and 
$\forall s \in \mathcal{T}(B).$ $\exists t \in \mathcal{T}(A).\ t{\ll}s$.
\end{definition}
\begin{definition}
Protocol A is a \textit{specialized-restriction} of protocol B if 
$\forall t \in \mathcal{T}(A).$ $\exists s \in \mathcal{T}(B).\ t{\ll}s$. 
\end{definition}

Notice that all those relationships can be easily discovered by straightforward algorithms 
supported by OWL-DL reasoners. Those reasoners deal with the ontology descriptions and rule engines that consider our semantic rules for generating protocol traces. 

\begin{figure}
	\centering
	\includegraphics[width=4.5in]{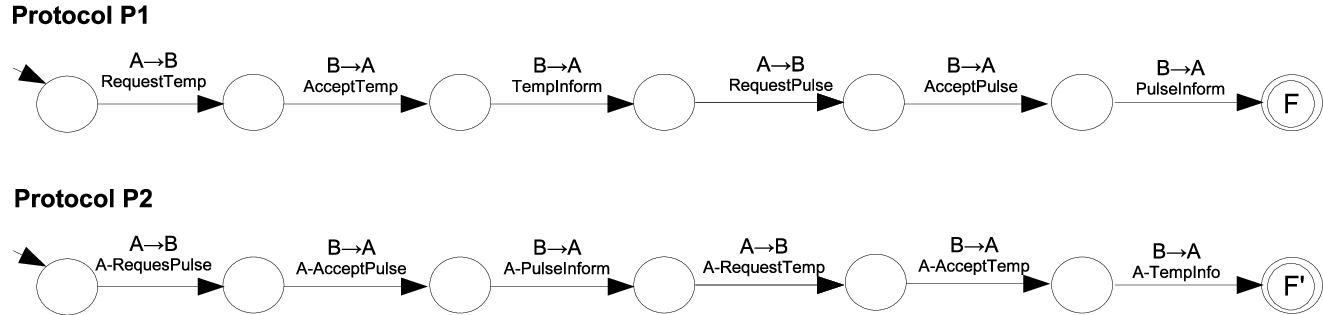}
	\caption{Specialization of protocols.} \label{ejemplo}
\end{figure}

Moreover, sometimes we may be interested in comparing protocol traces independently of 
time stamps. That is, we may be interested in knowing if a protocol produces the same 
fluents as another, in whatever order.
For example, in Fig.~\ref{ejemplo} we show two protocols that can be used to ask for 
information related to the vital signs temperature and pulse. In fact, for that purpose 
it is irrelevant the order in which the two requests are done.

In protocol P1 we can find a general protocol, in which agent A makes a request about the temperature using the communication act \texttt{RequestTemp} for that purpose. Then, agent B 
accepts and replies with a \texttt{TempInform} message, which is used to give information about the temperature. Once agent A receives this information, it asks agent B information about the pulse using a \texttt{RequestPulse}. Finally agent B accepts and replies with a \texttt{PulseInform} message and the final state F is reached. On the other hand, in protocol P2 we can find the specific protocol used by the agents of a specific system, called \textsc{Aingeru}\footnote{The A- prefix intends to label the \textsc{AINGERU} terminology}, to exchange information about vital signs. Protocol P2 may be a specialization of an standard protocol. First, agent A asks for the pulse, using the communication act \texttt{A-RequestPulse}. Then, agent B accepts and responds to the request using the \texttt{A-PulseInform} message. Next, agent A sends a \texttt{A-RequestTemp} message to ask about the temperature. Finally, agent B accepts and replies using the \texttt{A-TempInform} message and reaches state F'. 
Following we show the OWL specification for the communication acts used in this example.

\scriptsize
\begin{eqnarray*}
\texttt{RequestTemp} & \equiv & \texttt{Request} \sqcap \texttt{=1 hasContent.TempReq}\\
\texttt{AcceptTemp} & \equiv & \texttt{Accept} \sqcap \texttt{=1 hasContent.TempReq}\\
\texttt{TempInform} & \equiv & \texttt{Responsive} \sqcap \texttt{=1 hasContent.TempInfo} \sqcap \texttt{=1 inReplyTo.RequestTemp}\\
\texttt{RequestPulse} & \equiv & \texttt{Request} \sqcap \texttt{=1 hasContent.PulseReq}\\
\texttt{AcceptPulse} & \equiv & \texttt{Accept} \sqcap \texttt{=1 hasContent.PulseReq}\\
\texttt{PulseInform} & \equiv & \texttt{Responsive} \sqcap \texttt{=1 hasContent.PulseInfo} \sqcap \texttt{=1 inReplyTo.RequestPulse}\\
\texttt{A-RequestTemp} & \equiv & \texttt{RequestTemp} \sqcap \texttt{=1 theSystem.Aingeru} \sqcap \texttt{=1 hasContent.A-TempReq} \\
\texttt{A-AcceptTemp} & \equiv & \texttt{AcceptTemp} \sqcap \texttt{=1 theSystem.Aingeru} \sqcap \texttt{=1 hasContent.A-TempReq} \\
\texttt{A-TempInform} & \equiv & \texttt{TempInform} \sqcap \texttt{=1 theSystem.Aingeru} \sqcap \texttt{=1 hasContent.A-TempInfo} \sqcap \\ & & \texttt{=1 inReplyTo.A-RequestTemp}\\
\texttt{A-RequestPulse} & \equiv & \texttt{RequestPulse} \sqcap \texttt{=1 theSystem.Aingeru} \sqcap \texttt{=1 hasContent.A-PulseReq} \\
\texttt{A-AcceptPulse} & \equiv & \texttt{AcceptPulse} \sqcap \texttt{=1 theSystem.Aingeru} \sqcap \texttt{=1 hasContent.A-PulseReq} \\
\texttt{A-PulseInform} & \equiv & \texttt{PulseInform} \sqcap \texttt{=1 theSystem.Aingeru} \sqcap \texttt{=1 hasContent.A-PulseInfo} \sqcap \\ & & \texttt{=1 inReplyTo.A-RequestPulse}
\end{eqnarray*}
\normalsize

Notice that every communication act in protocol P2 is a subclass of its counterpart 
in protocol P1 (i.e. $\texttt{A-RequestPulse} \sqsubseteq  \texttt{RequestPulse}$, etc.) 
and correspondingly $\texttt{A-PulseInfo} \sqsubseteq  \texttt{PulseInfo}$, etc., is also 
satisfied.

Through a reasoning procedure analogous to that explained with the example of the AskTime 
protocol, we get the following sets of protocol traces: 

$\mathcal{T}(P1) = \{$[(\textit{accept(B, A, TempReq)}, $t_{2}$), (\textit{TempInfo}, $t_{4}$), (\textit{accept(B, A,}

\noindent\textit{PulseReq)}, $t_{6}$), (\textit{PulseInfo}, $t_{8}$)] $\}$ 

$\mathcal{T}(P2) = \{$[(\textit{accept(B, A, A-PulseReq)}, $t_{2}$), (\textit{A-PulseInfo}, $t_{4}$), (\textit{accept(B, A, A-TempReq)}, $t_{6}$), (\textit{A-TempInfo}, $t_{8}$)] $\}$

Even if the structure of the protocols is not exactly the same,
we can relate both protocols by a shallow notion of specialization from the following 
point of view. If we get abstracted  from time stamps, we can see protocol traces as multi-sets. 
Let us denote \textit{abstract-time(t)} to the multi-set formed by the fluents appearing in 
the protocol trace $t$, without any time stamp associated. Now, we define

  $\mathcal{S}(A) = \{ \textit{abstract-time(t)}  |  t \in \mathcal{T}(A)\}$ 

Then, we are in condition to define analogous relationships to the previous five, but in a 
shallow mood.
\begin{definition}

\begin{enumerate}
\item 
Protocol A is \textit{shallow-equivalent} to protocol B if $\mathcal{S}(A) = \mathcal{S}(B)$.
\item
Protocol A is a \textit{shallow-restriction} of protocol B if $\mathcal{S}(A) \subset  \mathcal{S}(B)$.
\item
A protocol trace $t$ is a \textit{shallow-specialization} of a protocol trace $s$,  
written $t\ll_{s}s$, if there is a map $\phi$ from \textit{abstract-time(t)} to 
\textit{abstract-time(s)} 
such that $\forall f\in \textit{abstract-time(t)}. f\sqsubseteq \phi(f)$ in an ontology of fluents.
\item
Protocol A is a \textit{shallow-specialized-equivalent} of protocol B if 
$\forall t \in \mathcal{S}(A).\ \exists s \in \mathcal{S}(B).\ t\ll_{s}s$ and 
$\forall s \in \mathcal{S}(B).\ \exists t \in \mathcal{S}(A).\ t\ll_{s}s$.
\item
Protocol A is a \textit{shallow-specialized-restriction} of protocol B if 
$\forall t \in \mathcal{S}(A).\ \exists s \in \mathcal{S}(B).\ t\ll_{s}s$. 
\end{enumerate}
\end{definition}

Finally, using our proposal, we can conclude that protocols P1 and P2 are  
\textit{shallow-specialized-equivalent}, although they use different communications 
acts and have different structure.

\section{Related Works}\label{trabajos} 
Among the different related works that we can find in the specialized literature, 
the closer work is \cite{MallyaSingh2007}, where protocols are represented as transition 
systems and 
subsumption and equivalence of protocols are defined with respect to three state similarity 
funtions. We share some goals with that work, but the protocol description formalism used by them is
not considered in the paper and there is no references to how protocol relationships are 
computed. In contrast, we describe protocols with a description logic language and 
protocol relationships can be computed by straightforward algorithms. It is worth 
mentioning that protocol relationships considered in that paper deserve study in our 
framework. 

 The works of \cite{YolumSingh2002} and \cite{Fornara03} are quite similar one to 
each other. Both capture the semantics of communication acts 
through agents' commitments and represent communication protocols using a set of rules that 
operate on these commitments. Moreover those rule sets can be compiled as finite state machines. Nevertheless, they do not consider the study of relationships between protocols. In addition, in \cite{DesaiMallyaEtAl2005}, protocols are also represented with a set of rules with terms 
obtained from an ontology, but their main goal is protocol development and, in order to 
reason about protocol composition, they formalize protocols into the $\pi$-calculus. 
Then, equivalence through bisimulation is the only process relationship considered. 
 In \cite{YS07}, they also consider commitment protocols; however, their main focus is on combining them with considerations of rationality on the enactment of protocols. Our proposal could be complemented with their approach.

An alternative way to describe finite state machines with a description logic language 
is to take advantage of the relationship of that logic with Deterministic Propositional 
Dynamic Logic, see \cite{BerardiEtAl2005} for an example in the context of Web Services 
composition. The approach of that paper is very different in purpose from ours. 
Their states and transitions descriptions are not prepared to be confronted 
in a comparison. In constrast, our state and transition descriptions are carefully modelled  
as class descriptions such that semantics relationships between protocols can be captured.

Also in the context of Web Services, state transition systems are used in 
\cite{BordeauxBerardiEtAl2004} for representing dynamic behaviour of services and 
they  define some notions 
of compatibility and substitutability of services that can be easily translated to 
the context of compatibility of protocols. Relationships between their compatibility 
relations and our defined relationships deserve study.

In \cite{KagalFinin2007} protocols are defined as a set of permissions and 
obligations of agents participating in the communication. They use an OWL ontology 
for defining the terms of the specification language, but their basic reasoning is 
made with an ad hoc reasoning engine. We share their main goal of defining protocols 
in a general framework that allows reutilization. Nevertheless, they do not consider 
relationships between protocols. 

The problem of determining if an agent's policy is conformant to a protocol is 
a very important one, but we are not treating that topic in this paper. 
Nevertheless, the topic is close to ours and it is worth mentioning the following papers 
that consider different notions of conformance:   
In \cite{Endriss03}, deterministic finite state machines are the abstract models for 
protocols, which are described by simple logic-based programs. Three levels of 
conformance are defined: weak, exhaustive and robust.
They consider communication acts as atomic actions, in contrast to our semantic view. 
In \cite{Baldoni06-ICSOC}a nondeterministic finite state automata is used to support a 
notion of conformance that guarantees interoperabiliy among agents conformant to a 
protocol. Their conformance notion considers the branching structure of policies and 
protocols and applies a simulation-based test. Communication acts are considered 
atomic actions, without considering their semantics.
In \cite{Chopra-Singh-06}, communication acts semantics is described in terms of 
commitments but it is not used for the conformance notion. A third different notion 
of conformance is defined and, moreover, it is proved orthogonal to their proposed 
notions of coverage and interoperability.

Finally, \cite{D'inverno98} and \cite{Mazouzi02} use 
finite state machines and Petri nets, respectively, but without taking into account 
the meaning of the communication acts interchanged, neither considering relationships 
between protocols.
  
\section{Conclusions}
Increasing machine-processable tasks in the Web is a challenge considered at present. 
In this line we have presented in this paper a proposal that favours the communication among 
agents that represent to different Information Systems 
accessible through the Web. The main contributions of the proposal are:
\begin{itemize}
\item The management of the semantics aspects when dealing with agent communication protocols.
\item The provision of the possibility of customizing standard communication protocols and management of them.
\item The use of standard Semantic Web tools to describe protocols.
\item The support for discovering different kinds of relationships between protocols.

\end{itemize}

\bibliographystyle{splncs}
\bibliography{BDI}
\end{document}